\documentclass{mn2e}
\usepackage{epsfig,natbib2,natbibmnfix}
\usepackage{amssymb}
\usepackage{multirow}
\usepackage{multicol}
\usepackage[varg]{txfonts}
\usepackage{color}

\newcommand{\fref}[1]{Fig.~\ref{#1}}

\newcommand{\eref}[1]{equation~\ref{#1}}
\newcommand{\cref}[1]{Chapter~\ref{#1}}

\newcommand{\note}[1]{\textcolor{red}{#1}}

\def \mathbi#1{\textbf{\em #1}}

\begin{document}
\title[Disc--binary alignment]{The final parsec problem: aligning a
  binary with an external accretion disc}

\author[C.J.~Nixon, A.R.~King \& J.E.~Pringle] 
{
\parbox{5in}{C.J.~Nixon$^1$, A.R.~King$^1$ and J.E.~Pringle$^{1,2}$}
\vspace{0.1in} 
 \\ $^1$ Theoretical Astrophysics Group, University of Leicester,
 Leicester LE1 7RH UK  
 \\ $^2$ Institute of Astronomy, University of Cambridge, Madingley Road,
 Cambridge CB3 0HA 
}

\maketitle

\begin{abstract}
We consider the interaction between a binary system (e.g. two
supermassive black holes or two stars) and an external accretion disc
with misaligned angular momentum. This situation occurs in galaxy
merger events involving supermassive black holes, and in the formation
of stellar--mass binaries in star clusters. We work out the
gravitational torque between the binary and disc, and show that their
angular momenta $\mathbi{J}_{\rm{b}}, \mathbi{J}_{\rm{d}}$ stably
counteralign if their initial orientation is sufficiently retrograde,
specifically if the angle $\theta$ between them obeys $\cos\theta <
-J_{\rm{d}}/2J_{\rm{b}}$, on a time short compared with the mass gain
time of the central accretor(s). The magnitude $J_{\rm{b}}$ remains
unchanged in this process. Counteralignment can promote the rapid
merger of supermassive black hole binaries, and possibly the formation
of coplanar but retrograde planets around stars in binary systems.
\end{abstract}


\footnotetext[1]{E-mail: Chris.Nixon@astro.le.ac.uk}

\section{Introduction}
\label{intro}

Galaxy mergers are commonly thought to be the main mechanism driving the
coevolution of galaxies and their central supermassive black holes (SMBH). In
such a merger we expect the formation of a SMBH binary in the centre of the
merged galaxy. Gravitational waves quickly drive the binary to coalesce if the
orbital separation can be shortened to $\lesssim 10^{-2}$pc. The binary may
stall at a separation greater than this if the interaction with the merged
galaxy is not efficient enough in extracting orbital angular momentum and
energy. For the stellar component of the galaxies this occurs at approximately
a parsec, creating ``the final parsec problem'' \citep{MM2001}. There have
been many papers exploring potential solutions to this problem, for example a
sling-shot mechanism involving a triple SMBH system \citep{Iwasawaetal2006},
efficient refilling of the binary loss cone by angular momentum exchange
between stellar orbits and a triaxial dark matter halo \citep{Bercziketal2006}
and also the evolution of the binary with a prograde accretion disc
(circumbinary discs: \citealt{AN2005}; \citealt{MM2008};
\citealt{Lodatoetal2009}; \citealt{Cuadraetal2009} and embedded discs:
\citealt{Escalaetal2005}; \citealt{Dottietal2007,Dottietal2009}). In a recent
paper \citep{Nixonetal2011} we explored the evolution of a binary interacting
with a {\it retrograde} circumbinary accretion disc. We showed that this is
more efficient than a prograde disc in removing binary orbital angular
momentum and energy. This is simply because there are no orbital resonances
between the binary and the disc and thus there is direct accretion of
retrograde gas onto the binary.

Here we consider the alignment process between a binary system and an external
misaligned accretion disc. This situation can arise in at least two
astronomical contexts. First, a merger event between galaxies can produce a
SMBH binary in the centre of the merged galaxy, and this or a later accretion
event may surround the hole with a disc of accreting gas. A similar situation
arises during the formation of stars in a cluster. A binary system may form,
but also capture gas into an external disc.

In both of these cases, there is no compelling reason to assume that the
binary and disc rotation are initially parallel or even roughly coaligned (cf
\citealt{KP2006}). As we shall see, the gravitational interaction between the
binary and the disc generates differential precession in the disc gas, and
thus viscous dissipation. This gives a dissipative torque which vanishes only
when the binary and disc angular momenta $\mathbi{J}_{\rm{b}},
\mathbi{J}_{\rm{d}}$ are either parallel or antiparallel. In all such cases,
the torque diffuses the tilt or warp through the disc (cf
\citealt{Pringle1992,Pringle1999}; \citealt{WP1999}) driving the system to one
of these equilibria. The existence of a warp makes the precise definition of
disc angular momentum $\mathbi{J}_{\rm{d}}$ quite subtle and we return to this
point in the Discussion.

The binary--external disc interaction is very similar to the effect of the
Lense--Thirring (LT) precession on an accretion disc around a spinning black
hole (\citealt{BP1975}; \citealt{Pringle1992}; \citealt{SF1996};
\citealt{NP1998}; \citealt{AN1999}; \citealt{NA1999}; \citealt{NP2000};
\citealt{LP2006} etc) if we replace $\mathbi{J}_{\rm{b}}$ by the hole spin
angular momentum $\mathbi{J}_{\rm{h}}$. For some years it was thought that the
LT interaction always led to co--alignment (i.e. $\mathbi{J}_{\rm{b}}$ and
$\mathbi{J}_{\rm{d}}$ parallel). However \citet{Kingetal2005} (hereafter KLOP)
showed on very general grounds that counteralignment does occur, if (and only
if) the initial angle $\theta$ between $\mathbi{J}_{\rm{d}}$ and
$\mathbi{J}_{\rm{h}}$ satisfies $\cos\theta < -J_{\rm{d}}/2J_{\rm{h}}$, where
$J_{\rm{d}} = \left|\mathbi{J}_{\rm{d}} \right|$ and $J_{\rm{h}} =
\left|\mathbi{J}_{\rm{h}} \right|$. \citet{SF1996} had implicitly assumed
$J_{\rm{d}} \gg J_{\rm{h}}$ and so enforced co--alignment. With this
restriction lifted, \citet{KP2006,KP2007} and \citet{Kingetal2008} showed that
accretion from a succession of randomly--oriented discs leads to spindown of
the supermassive black hole, allowing rapid mass growth.

In this paper we examine the alignment process for a binary and an external
disc. We show that the argument of KLOP is generic, and that the disc and
binary counteralign if and only if $\cos\theta < -J_{\rm{d}}/2J_{\rm{b}}$. As
a result it is quite possible for SMBH binaries to be surrounded by a
completely retrograde disc which strongly promotes coalescence (cf
\citealt{Nixonetal2011}). In the case of a newly--formed stellar binary, the
presence of a counteraligned disc can lead to the formation of planets with
retrograde orbits.

\section{The binary--disc torque}
\label{binarydisctorque}
We consider a binary system with masses $M_1, M_2$ and a circular orbit, with
the binary angular momentum vector pointing along the $z$--axis of cylindrical
polar coordinates $(R, \phi, z)$. For simplicity we assume $M_2 \ll M_1$ and
place $M_1$ at the origin, with $M_2$ orbiting at radius $a$ in the $(R,
\phi)$ plane (our conclusions are not affected by this assumption). The orbit
has angular velocity
\begin{equation}
\Omega_{\rm{b}} = \biggl[{G(M_1+M_2)\over a}\biggr]^{1/2}.
\end{equation}

Now we consider a disc particle in an orbit about the binary at radius $R \gg
a$. If both the small quantities $M_2/M_1$ and $a/R$ actually vanished, the
particle's orbit would be a circle, with angular velocity
$(GM_1/R)^{1/2}$. When these quantities are small but finite they induce
various perturbations in the orbit. Some of these perturbations have
(inertial--frame) frequency $2\Omega_{\rm{b}}$ and higher multiples. These are
oscillatory, and have no long--term secular effect. Long--term effects on the
orbit, and hence eventually on the disc, come from the zero--frequency
(azimuthally symmetric $m = 0$) term in the binary potential. This point is
considered in more detail in \citet{Bateetal2000}, who considered the related
problem of a disc around the primary mass $M_1$ (i.e. $M_2 \ll M_1$, but $R
\ll a$).

Physically this $m=0$ term is given by replacing the orbiting mass $M_2$ with
the same mass spread uniformly over its orbit, i.e. a ring of mass $M_2$ and
radius $a$ in the $(R, \phi)$ plane. Adding in the potential from the fixed
point mass $M_1$ at the origin we find the effective gravitational potential
experienced by a disc particle as

\begin{equation}
\Phi(R, z) = -{GM_1\over (R^2 + z^2)^{1/2}} - {GM_2\over
  2\pi}\int_0^{2\pi}{{\rm d}\phi\over r}
\label{Phi}
\end{equation}
where $r$ is the distance between the particle position and a point on
the ring at $(a, \phi, z)$, i.e.
\begin{equation}
r^2 = R^2 + a^2 + z^2 - 2Ra\cos\phi.
\end{equation}
We now expand \eref{Phi} in powers of $a/R$ and $z/R$, keeping
terms only up to second order. This gives
\begin{eqnarray}
\lefteqn{\Phi(R, z) = -{G(M_1 + M_2)\over R} + {GM_2a^2\over 4R^3} +
  {G(M_1+M_2)z^2\over 2R^3}} \nonumber \\ 
\lefteqn{\ \ \ \ \ \ \ \ \ \ \ \ \ \ \ \ \ - {9GM_2a^2z^2\over
    8R^5} + ....}
\label{Phi2}
\end{eqnarray}
The orbital frequency $\Omega$ of the particle subject to this
potential is given by
\begin{equation}
\Omega^2 = {1\over R}{\partial\Phi\over \partial R}
\end{equation}
and its vertical oscillation frequency $\nu$ by
\begin{equation}
\nu^2 = {\partial^2\Phi\over \partial z^2}, 
\end{equation} 
both evaluated at $z=0$.
The nodal precession frequency is $\Omega_p = \Omega - \nu$ and
we find
\begin{equation}
\Omega_{\rm{p}}(R) = {3\over 4}\biggl[{G(M_1+M_2)\over
  R^3}\biggr]^{1/2}{M_2\over M_1+M_2}{a^2\over R^2}.
\label{prec}
\end{equation}

This frequency is very similar to that for LT precession around a spinning
black hole (e.g. \citealt{SF1996}), which goes as $R^{-3}$ rather than the
$R^{-7/2}$ here.  Equation~\ref{prec} is formally almost identical to the
precession frequency found by \citet{Bateetal2000} for a disc around the
primary, although derived for $R \gg a$ rather than $R \ll a$. The same
argument as in that paper shows that if the disc and binary axis are
misaligned by an angle $\theta$ (called $\delta$ in \citealt{Bateetal2000})
with $0\leq \theta \leq \pi/2$, the precession frequency is just multiplied by
$\cos\theta$. The opposite case with the disc somewhat counteraligned
(i.e. $\theta > \pi/2$) is equivalent to the $\theta < \pi/2$ case with the
binary angular momentum reversed. But this reversal leaves the precession
frequency unchanged, since we are dealing only with the $m=0$ part of the
potential. So for all $\theta$ with $0 < \theta < \pi$ the precession
frequency is
\begin{equation}
\Omega_{\rm{p}}(\theta) = \Omega_{\rm{p}} \left| \cos\theta \right|.
\end{equation}
This result differs from the LT case, where the factor
$\cos\theta$ appears without modulus signs.

\section{Co-- or Counter--alignment?}
\label{align_coalign}
We have shown above that the effect of the binary potential on the disc is to
induce precession of the disc orbits. This precession is strongly dependent on
radius: rings of gas closer to the binary precess faster. The differential
precession creates a dissipative torque between adjacent rings of gas tending
to make $\theta \longrightarrow 0, \pi$ so that the precession ultimately
vanishes.

The precession timescale in the disc increases with radius
(cf.~\eref{prec}). The torque therefore acts faster at smaller radii to co--
or counteralign disc orbits with the binary plane. This leads to the creation
of a warp in the disc, where the inner parts are co-- or counteraligned and
the outer parts are still misaligned (cf.~\fref{disc}). This warp propagates
outwards, eventually co-- or counter aligning the entire disc with the binary
plane. This effect was solved numerically for discs warped under the LT effect
by \citet{LP2006}.


Now we argue as in KLOP that since each ring feels a precession, the resultant
back--reaction on the binary is a sum of precessions, which is just a
precession. This argument is equivalent to that presented in
\citet{Bateetal2000} who argue that because the binary potential is symmetric
about the plane of the binary, the disc--binary torque cannot have a component
in the direction of $\mathbi{J}_{\rm{b}}$. Accordingly $\mathbi{J}_{\rm{b}}$
can only precess. These arguments show that we can write the torque on the
binary in the same form as the LT--induced torque on a spinning black hole
considered in KLOP, i.e.
\begin{equation}
\frac{{\rm d} \mathbi{J}_{\rm{b}}}{{\rm d}t} = - K_1 [ \mathbi{J}_{\rm{b}} \wedge
  \mathbi{J}_{\rm{d}} ] - K_2 [ \mathbi{J}_{\rm{b}} \wedge (\mathbi{J}_{\rm{b}} \wedge
  \mathbi{J}_{\rm{d}}) ].
\label{align}
\end{equation}
Here $K_1, K_2$ are coefficients depending on disc properties. The first term
gives the magnitude and sign of the torque inducing the precession. It does
not change the alignment angle $\theta$. The second term describes the torque
which changes $\theta$. The same arguments as in KLOP for the LT case, and
\citet{Bateetal2000} for a disc around the primary, show that dissipation in
the disc requires $K_2$ to be a positive quantity. Its magnitude depends on
the properties of the disc and the binary. The one difference from the LT case
is that the $\left|\cos\theta \right|$ dependence means that the sign of the
coefficient $K_1$ can be either positive or negative. But this difference has
no effect on the conditions under which the disc and binary co-- or
counteralign. These are formally identical with the ones for the LT case
derived by KLOP, with the binary angular momentum $\mathbi{J}_{\rm{b}}$
replacing the hole spin angular momentum $\mathbi{J}_{\rm{h}}$. The process
obviously has a different timescale specified by the different magnitude of
$K_2$.

The same arguments as in KLOP now show that the magnitude $J_{\rm{b}}$ of the
binary angular momentum remains constant, while the direction of
$\mathbi{J}_{\rm{b}}$ aligns with the total angular momentum
$\mathbi{J}_{\rm{t}} = \mathbi{J}_{\rm{b}} + \mathbi{J}_{\rm{d}}$, which is of
course a constant vector. During this process the magnitude of $J_{\rm{d}}^2$
decreases because of dissipation (KLOP).  Counteralignment
($\theta \rightarrow \pi$) occurs if and only if $J_{\rm{b}}^2 >
J_{\rm{t}}^2$. By the cosine theorem
\begin{equation}
J_{\rm{t}}^2 = J_{\rm{b}}^2 + J_{\rm{d}}^2 -
2J_{\rm{b}}J_{\rm{d}}\cos(\pi - \theta),
\label{cos}
\end{equation}
so this is equivalent to
\begin{equation}
\cos\theta < -{J_{\rm{d}}\over 2J_{\rm{b}}}.
\label{crit}
\end{equation}
Thus counteralignment of a binary and an external disc is possible, and
requires
\begin{equation}
\theta > \pi/2,\ \  J_{\rm{d}} < 2J_{\rm{b}}. 
\label{cond}
\end{equation}

\section{Discussion}
\label{discussion}

So far in this paper we have avoided fully spelling out the meaning of the
disc angular momentum $\mathbi{J}_{\rm{d}}$. This is complicated because the
binary torque falls off very strongly with radius, and so a large
contribution to the angular momentum in a distant part of the disc may be
irrelevant to the alignment process, or affect this process in a
time--dependent way (cf \citealt{LP2006}). Sections 3 and 4 of KLOP discuss
these questions in more detail. Effectively $\mathbi{J}_{\rm d}$ can be
thought of as the disc angular momentum inside the warp radius, and therefore
a time--dependent quantity. 

At early times $J_{\rm{d}}$ is small, as
only a fraction of the total gas interacts with the
binary. Counter--alignment may occur if $\theta > \pi/2$, but at later times,
as $J_{\rm{d}}$ grows and more gas is able to interact with the binary,
alignment eventually happens (when $J_{\rm{d}} > 2J_{\rm{b}}$). So if $\theta
>\pi/2$, even for $J_{\rm{d}} > 2J_{\rm{b}}$ we expect $\sim
2J_{\rm{b}}|\cos\theta|$ of disc angular momentum to counteralign with the
binary before the outer disc comes to dominate and enforce coalignment (cf. \citealt{LP2006}).

\begin{figure}
  \begin{center}
  \includegraphics[angle=0,width=20pc]{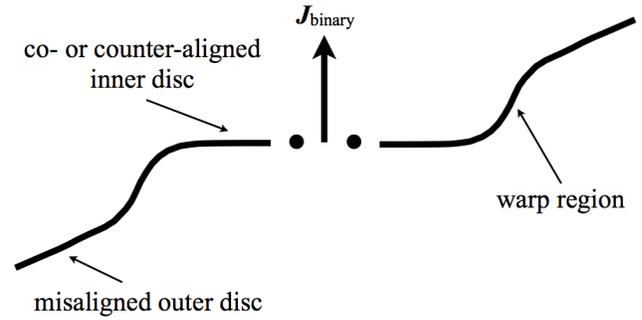}
  \caption{The warped disc shape expected after the inner disc co-- or
    counter--aligns with the binary plane but the outer disc stays
    misaligned. Eventually the entire disc will co--or counter--align with the
    binary plane, depending on the global criterion (eqn~\ref{cond}). Note
    that in practice precession makes the warp non--axisymmetric.}
  \label{disc}
  \end{center}
\end{figure}

The typical timescale for co-- or counter--alignment for a SMBH binary is
\begin{equation}
t_{\rm binary} \simeq {J_{\rm{b}}\over
  J_{\rm{d}}(R_{\rm{w}})}{R_{\rm{w}}^2\over \nu_2}
\label{talign}
\end{equation}
where $R_{\rm{w}}$ is the warp radius, $J_{\rm{d}}(R_{\rm{w}})$ is the disc
angular momentum within $R_{\rm{w}}$, and $\nu_2$ is the vertical disc
viscosity. This is identical to the formal expression for LT alignment of a
spinning black hole if we replace the spin angular momentum $J_{\rm{h}}$ with
$J_{\rm{b}}$ (cf \citealt{SF1996}). The warp radius is given by equating the
precession time $1/\Omega_{\rm{p}}(R)$ to the vertical viscous time
$R^2/\nu_2$ . Inside this radius the precession timescale is short and the
disc dissipates and co-- or counter--aligns with the binary plane. Outside
this radius the disc is not dominated by the precession and so maintains its
misaligned plane. The connecting region therefore takes on a warped shape
shown in \fref{disc}. As time passes the warp propagates outwards and
co-- or counter--aligns the entire disc with the binary plane.

Approximating the disc angular momentum as
\begin{equation}
J_{\rm{d}}(R_{\rm{w}}) \sim \pi R_{\rm{w}}^2\Sigma(GMR_{\rm{w}})^{1/2}
\label{jwarp}
\end{equation}
with $\Sigma$ the disc surface density and $M = M_1 + M_2$, and using
the steady--state disc relation $\dot M = 3\pi\nu\Sigma$ we find
\begin{equation}
t_{\rm binary} \sim 3{M_2\over M_1}\biggl({a\over
  R_{\rm{w}}}\biggr)^{1/2}{\nu_1\over \nu_2}{M\over \dot M},
\label{talign2}
\end{equation}
where we have also used 
\begin{equation}
J_{\rm{b}} = M_1M_2\biggl({Ga\over M}\biggr)^{1/2}.
\label{jb}
\end{equation}
Since $\nu_1 < \nu_2$ \citep{PP1983}, $a \ll R_{\rm{w}}$ and $M_2 < M_1$, we
see that alignment takes place on a timescale shorter than the mass growth of
the central accretor(s).

The timescale (\ref{talign2}) is directly analogous to the expression
\begin{equation}
t_{\rm LT} \sim 3a_*\biggl({R_{\rm{s}}\over
  R_{\rm{w}}}\biggr)^{1/2}{\nu_1\over \nu_2}{M\over \dot M}
\label{lt}
\end{equation}
for alignment under the LT precession, where $a_* < 1$ is the Kerr spin
parameter and $R_{\rm{s}}$ the Schwarzschild radius of the spinning
hole. Evaluating $R_{\rm{w}}$ in the two cases we find
\begin{equation}
{t_{\rm LT}\over t_{\rm binary}} \sim {3^{1/2}\over 2}\biggl({a_*\over
  M_2/M_1}\biggr)^{1/2}\biggl({a\over R_{\rm{s}}}\biggr)^{1/4}.
\end{equation}
Thus in general, provided we assume that the ratio $\nu_1/\nu_2$ is similar in
the two cases and that the hole spin is not rather small ($a_* <
(R_{\rm{s}}/a)^{1/2}(M_2/M_1)$), then the binary--disc alignment is rather
faster than the corresponding process for spinning black holes.

Our result has significant consequences for SMBH binaries. For random
orientations, \eref{cond} shows that initial disc angles leading to
alignment occur significantly more frequently than those giving
counteralignment only if $J_{\rm d} > 2J_{\rm b}$. (In the LT case this fact
leads to a slow spindown of the hole, because retrograde accretion has a
larger effect on the spin, \citealt{Kingetal2008}.) A number of studies
(\citealt{AN2005}; \citealt{MM2008}; \citealt{Cuadraetal2009};
\citealt{Lodatoetal2009}) have shown that prograde external discs are rather
inefficient in shrinking SMBH binaries and solving the last parsec
problem. This is essentially because of resonances within the disc. In
contrast, the slightly rarer retrograde events have a much stronger effect on
the binary. These rapidly produce a counterrotating but coplanar accretion
disc external to the binary, which has no resonances. We note that
\citet{Nixonetal2011} show that the binary gradually increases its
eccentricity as it captures negative angular momentum from the disc,
ultimately coalescing once this cancels its own. A non--zero binary
eccentricity changes the detailed form of the perturbing potential from that
in \eref{Phi2}, but cannot change the precessional character leading to the
torque equation (\ref{align}). Our results remain unchanged, particularly the
counteralignment condition (\ref{cond}), apart from minor modifications of the
timescale (\ref{talign2}).

Thus in a random sequence of accretion events producing external discs, the
prograde events have little effect, and the retrograde ones shrink the
binary. In particular, a sequence of minor retrograde events with $J_{\rm{d}}
< J_{\rm{b}}$ has a cumulative effect and must ultimately cause the binary to
coalesce once the total retrograde $\sum J_{\rm{d}} = J_{\rm{b}}$.  This is
important, since the disc mass is limited by the onset of self--gravity to
$M_{\mathrm{d}} \lesssim \left(H/R \right) M_{1}$ (cf King, Pringle and
Hofmann 2008). Coalescence will then occur once the retrograde discs have
brought in a total mass $M_2$, i.e. once a sequence of $\gtrsim
(M_2/M_1)(R/H)$ retrograde discs have accreted. For minor mergers this
requires at most a few randomly oriented accretion disc events, rising
to a few hundred for major mergers ($q >0.1$).

We note finally that similar considerations apply in planet--forming discs
around stellar binary systems, which can also be initially misaligned
\citep{Bateetal2010}. This may offer a way of making retrograde planets in
binaries, as recently suggested for $\nu$ Octantis \citep{EC2010}.

\section*{Acknowledgments}
\label{acknowledgments}
We would like to thank the anonymous referee for helping to improve the
clarity of this work. CJN holds an STFC postgraduate studentship. Research in
theoretical astrophysics at Leicester is supported by an STFC Rolling Grant.

\bibliographystyle{mn2e} 
\bibliography{nixon}

\end{document}